# Soliton Radiation Beat Analysis of Optical Pulses Generated from Two CW Lasers

M. Zajnulina[a)], M. Böhm[b)], K. Blow[c)], A. A. Rieznik[d)], D. Giannone[a)], R. Haynes[a)], M. M. Roth[a)]

[a]innoFSPEC-VKS, Leibniz Institute for Astrophysics, An der Sternwarte 16, 14482 Potsdam, Germany
[b]innoFSPEC-InFaSe, University of Potsdam, Am Mühlenberg 3, 14476 Golm, Germany
[c]Aston Institute of Photonic Technologies, Aston Triangle, Birmingham, B4 7ET, United Kingdom
[d]Instituto Tecnologico de Buenos Aires and CONICET, Buenos Aires, Argentina

We propose a fibre-based approach for generation of optical frequency combs (OFC) with the aim of calibration of astronomical spectrographs in the low and medium-resolution range. This approach includes two steps: in the first step, an appropriate state of optical pulses is generated and subsequently moulded in the second step delivering the desired OFC. More precisely, the first step is realised by injection of two continuous-wave (CW) lasers into a conventional single-mode fibre, whereas the second step generates a broad OFC by using the optical solitons generated in step one as initial condition. We investigate the conversion of a bichromatic input wave produced by two initial CW lasers into a train of optical solitons which happens in the fibre used as step one. Especially, we are interested in the soliton content of the pulses created in this fibre. For that, we study different initial conditions (a single cosine-hump, an Akhmediev breather, and a deeply modulated bichromatic wave) by means of Soliton Radiation Beat Analysis and compare the results to draw conclusion about the soliton content of the state generated in the first step. In case of a deeply modulated bichromatic wave, we observed the formation of a collective soliton crystal for low input powers and the appearance of separated solitons for high input powers. An intermediate state showing the features of both, the soliton crystal and the separated solitons, turned out to be most suitable for the generation of OFC for the purpose of calibration of astronomical spectrographs.

**Optical frequency combs constitute a discrete optical spectrum with lines that are equidistantly positioned. Frequency combs generated in mode-locked lasers have been proposed and already successfully tested as calibration sources for high-resolution astronomical spectrographs. However, there is a variety of novel astronomical instruments that operate in the low- and medium resolution range. They also would profit from the deployment of frequency combs as calibration marks. We propose a fibre-based approach for optical frequency comb generation that is specifically suitable for spectrographs with low and medium resolution. This approach consists of two fibres fed with two continuous-wave lasers. To be able to generate broadband, stable, and low-noise frequency combs, we need to understand the optical pulse formation in different fibre stages. In this paper, we focus our attention on the pulse build-up in the first fibre stage of the proposed approach and study it numerically by means of the Soliton Radiation Beat Analysis.**

## 1. INTRODUCTION

Optical frequency combs (OFC) constitute an array of equidistantly-spaced spectral lines that have nearly equal intensity over a broad spectral range[1,2]. Combs generated in mode-locked lasers have been successfully demonstrated as calibration sources for astronomical spectrographs deployed for high-resolution spectroscopy[3-9]. In this resolution range, the comb line spacing typically varies from 1 GHz to 25 GHz[10,11]. Such novel instruments like PMAS at the Caral Alto Observatory 3.5 m Telescope, MUSE being developed for the Very Large Telescope (VLT) of the European Southern Observatory (ESO), and 4MOST being in development for the ESO VISTA 4.1 m Telescope operate, however, in the low- and medium resolution range and need OFC having line spacings going from slightly below 100 GHz to a few hundreds of GHz[12-14].

We propose an all-optical fibre-based approach for generation of OFC in the low- and medium resolution range with tuneable line spacing. It consists of two fibre stages where the first stage is a conventional single-mode fibre, the second one is a suitably pumped amplifying Erbium-doped fibre with anomalous dispersion. The initial input field comes from two continuous-wave lasers (CW) that generate a deeply modulated bichromatic cosine-wave[15-18].

The goal of this study is to understand and to control the temporal shape of the first fibre stage output. This control is important, because the temporal profile of pulses generated in the first stage will define the pulse shapes and, thus, the OFC build-up in the second fibre stage. To be able to generate broadband OFC, one should reduce the pulse duration in the second fibre stage as much as possible. Further, a perfect temporally periodic output after the first stage is needed to achieve a high level of the OFC lines sharpness (the temporal aperiodicity of the pulses within the train would lead to the broadening of the OFC lines and, thus, should be avoided). To fulfil this requirement, it is necessary to start with a periodic initial condition. A bichromatic deeply modulated cosinusoidal optical field generated by two CW lasers suits well such requirement. Moreover, it allows to control the OFC line spacing by tuning the laser frequency separation (LFS).

An effective pulse compression in the time domain is realised if some interfering optical solitons are exited in the first fibre stage. Additionally, the usage of solitons is helpful to stabilise the output structure of this stage. However, if too many solitons are excited, they will tend to break up and, so, impact the periodicity of the temporal shape.

The propagation behaviour of a single soliton is well known[19]. However, we are here interested in a state when many solitons strongly overlap with each other. The soliton overlap that occurs in our system it not sufficiently understood yet in the general case. This system is described by a nonintegrable propagation equation. To get insight into the nonlinear dynamics that take place in the first fibre stage, we apply the Soliton Radiation Beat Analysis (SRBA) that will help us to retrieve the soliton content since it is capable of dealing with nonintegrable equations with arbitrary initial conditions[20,21].

To decode the SRBA spectra, we analyse different initial conditions. More precisely, in addition to the desired bichromatic cosinusoidal input field, we also study a single cosine-hump since this will provide us with information about the behaviour when single solitons are well separated. To study the contrary case of strongly overlapping solitons, we decode the case of a maximally compressed Akmediev breather as initial condition.

The OFC line spacings of the proposed fibre-based approach coincide with the optical pulse repetition rates. Therefore, this approach can also be used for the generation of high-repetition ps-pulses for the component testing and optical sampling as well as in the ultra-high capacity transmission systems based on the optical time-division multiplexing in the field of telecommunication[18]. Indeed, several similar fibre-based approaches for the generation of ps-pulses for the telecommunication applications have been already reported in the past[22-28]. We believe that these approaches can also benefit from the results presented in this paper.

This paper is structured as follows: in Sec. 2, we present the experimental setup for generation of OFCs in fibres and the corresponding mathematical model, the concept of SRBA is depicted in Sec. 3 and the results are shown in Sec. 4, a conclusion is drawn in Sec. 5.

## 2. EXPERIMENTAL SETUP AND MATHEMATICAL MODEL

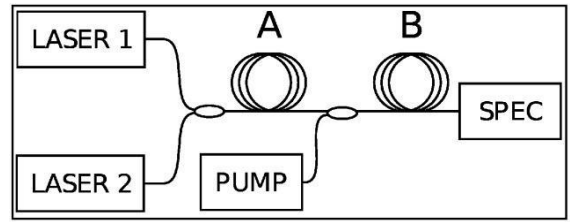

FIG. 1. Schematic presentation of the motivating setup. LASER1: fixed CW laser, LASER2: tuneable CW laser, A: conventional single-mode fibre, B: Er-doped fibre, PUMP: pump laser for fibre B, SPEC: astronomical spectrograph

The schematic representation of the experimental setup for generation of OFC in the low- and medium resolution range for the purpose of the spectrograph calibration is shown in Fig. 1[15-18]. The generation of an OFC starts with two independent and free-running CW lasers that have equal intensity and feature relative frequency stability of $10^{-8}$ over one-hour time frame sufficient for astronomical applications in the low- and medium-resolution range. Laser 1 is fixed at the angular frequency $\omega_1$, Laser 2 has the tuneable frequency $\omega_2$, the resulting modulated cosine-wave has the frequency $\omega_c = (\omega_1 + \omega_2)/2$ that coincides with the central wavelength at 1531 nm. The initial laser frequency separation, $LFS = |\omega_2 - \omega_1|/2\pi$, is $LSF = 78.125$ GHz. Fibre A is a conventional single-mode fibre, B is a pumped Erbium-doped fiber with anomalous dispersion. In fibre A, a sequence of temporally periodic soliton-like structures with widths in ps-range evolves out of the initial deeply modulated cosine-wave. These soliton-like structures have a narrowband OFC spectrum. In fibre B, the soliton-like pulses are compressed to the fs-range and the OFC broadened. Pulse compression in an amplifying medium can be considered as an alternative technique to the compression in dispersion-decreasing fibres[29-31].

We model the light propagation in fibre A by means of the generalised nonlinear Schrödinger equation (GNLS) for a slowly varying optical field envelope $A = A(z,t)$ in the co-moving frame[30,32,33]:

$$\frac{\partial A}{\partial z} = i \sum_{j=2}^{3} \frac{i^j}{j!} \beta_j \frac{\partial^j A}{\partial t^j} + i\gamma \left(1 + \frac{i}{\omega_0}\frac{\partial}{\partial t}\right) \times \\ \times \left(A \int_{-\infty}^{\infty} R(t')|A(t-t')|^2 dt'\right) \quad (1)$$

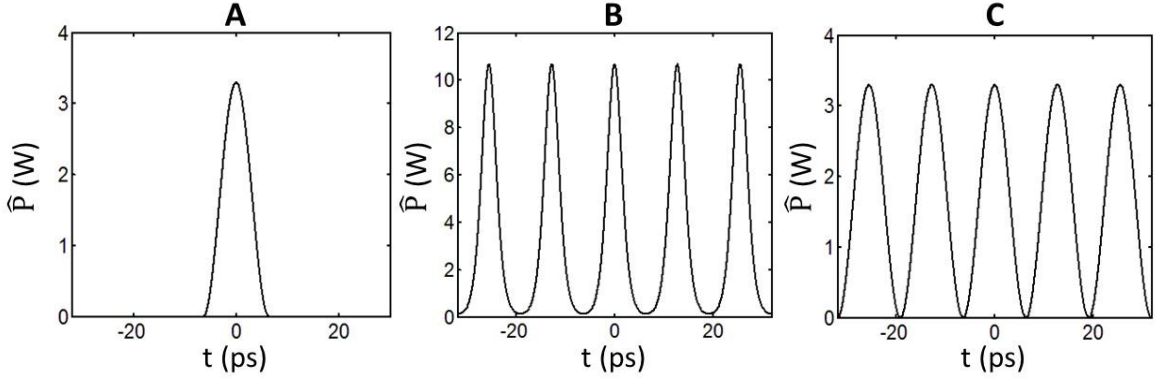

FIG. 2. Example of three types of initial conditions with initial power of $P_0 = 3.3$ W: A. Single cosine-hump, B. Maximally compressed Akhmediev breather, C. Deeply modulated cosine-wave according to the proposed setup

where $\beta_j = \left(\frac{\partial^j \beta}{\partial \omega^j}\right)_{\omega=\omega_0}$ is the value of the j$^{\text{th}}$ dispersion order at the carrier frequency $\omega_c$, whereas $\gamma = \frac{\omega_c n_2}{c \cdot S}$ defines the nonlinear coefficient with $n_2$ being the nonlinear refractive index of silica, $S$ the effective mode area, and $c$ the speed of light. The delayed Raman effect is incorporated via $h_R(t)$ into the response function that includes both, the electronic contribution assumed to be nearly instantaneous and the contribution set by vibration of silica molecules, and reads as:

$$R(t) = (1 - f_R)\delta(t) + f_R h_R(t) \quad (2)$$

with $f_R = 0.245$ denoting the fraction of the delayed Raman response to the total nonlinear polarisation. The function $h_R(t)$ is defined as:

$$h_R(t) = (1 - f_b)h_a(t) + f_b h_b(t),$$
$$h_a(t) = \frac{\tau_1^2 + \tau_2^2}{\tau_1 \tau_2^2} \exp\left(-\frac{t}{\tau_2}\right)\sin\left(\frac{t}{\tau_1}\right), \quad (3)$$
$$h_b(t) = \left(\frac{2\tau_b - t}{\tau_b^2}\right)\exp\left(-\frac{t}{\tau_b}\right)$$

where $\tau_1 = 12.2$ fs and $\tau_2 = 32$ fs are the characteristic times of the Raman response of silica and $f_b = 0.21$ represents the according vibrational instability with $\tau_b \approx 96$ fs[32,33].

For fibre A, following parameters are chosen: $\beta_2 = -15\frac{\text{ps}^2}{\text{km}}, \beta_3 = 0.1\frac{\text{ps}^3}{\text{km}}$, and $\gamma = 2\,\text{W}^{-1}\text{km}^{-1}$. The numerical solution of Eq. 1 is performed within a temporal window of 128 ps using the fourth-order Runge-Kutta in the interaction picture method in combination with the local error method with $2^{14}$ sample points[34].

The aim of this study is to investigate the soliton content of optical pulses generated using two CW lasers as presented in Fig. 1. To understand the pulse build-up in fibre A in detail, we first perform a gedankenexperiment: we consider the pulse build-up in two limit cases. The first limit case is given when a temporally localised structure is chosen as initial condition (IC). To be close to the initial condition of the proposed fibre-based approach for generation of OFC, we chose a single cosine-hump for the first limit (FIG. 2(A)):

$$A(z = 0, t) = \begin{cases} 0, & |t| > 6.4 \text{ ps} \\ N\sqrt{P_0}\cos(\omega_c t), & |t| \leq 6.4 \text{ ps} \end{cases} \quad (4)$$

Due to the properties of Eq. 1, the single cosine-hump is expected to evolve into a soliton with a sech-profile[19].

The other limit case is a temporally nonlocalised structure. We choose a maximally compressed Akhmediev breather, since Akhmediev breathers are well-known temporally periodic solutions of the Nonlinear Schrödinger Equation without additional terms (NLS) (FIG. 2(B))[35-38]:

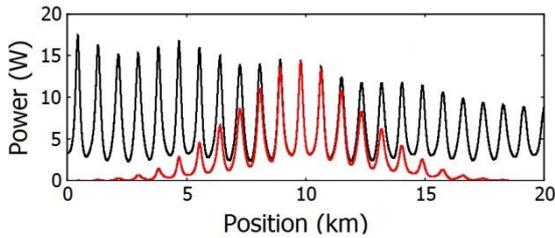

FIG. 3. Optical power at $t = 0$ (black) and apodized optical power at $t = 0$ (red) vs. propagation distance for $P_0 = 3.3$ W

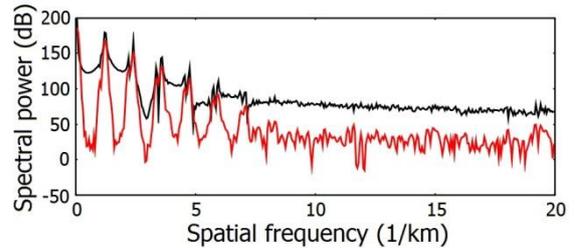

FIG. 4. Spectral power of non-apodized optical power (black) and apodized optical power (red) vs. spatial frequency (cf. FIG. 3)

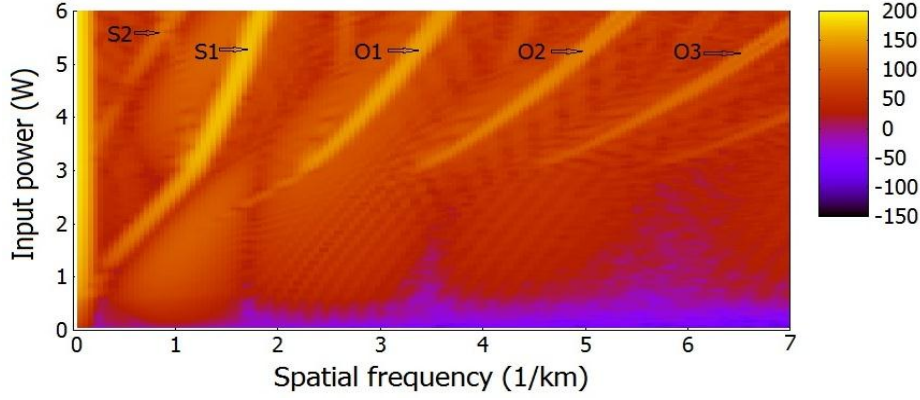

FIG. 5. Power spectrum for a single cosine-hump as initial condition for different values of the input power

$$A(z=0,t) = N\sqrt{P_0}\frac{(1-4a)+\sqrt{2a}\cos(\omega_{\mathrm{mod}}t)}{\sqrt{2a}\cos(\omega_{\mathrm{mod}}t)-1}, \quad (5)$$

where the coefficient $a$ is defined as $a = \frac{1}{2}\left(1-\frac{\omega_{\mathrm{mod}}^2|\beta_2|}{4\gamma P_0}\right)$ and the modulational frequency is $\omega_{\mathrm{mod}} = 4\pi LFS$.

The initial condition of the case we are actually interested in (radiation of two CW lasers) is described by (FIG. 2(C))[15-18]:

$$A(z=0,t) = N\sqrt{P_0}\cos(\omega_c t). \quad (6)$$

To increase the resemblance between the conditions presented in Eq. 5 and Eq. 6, we choose $a = 0.08$. In Eq. 4, Eq. 5, and Eq. 6, $N$ is the scale soliton order given by

$$N^2 = \frac{\gamma P_0}{(2\pi \cdot LFS)^2|\beta_2|}. \quad (7)$$

## 3. SOLITON RADIATION BEAT ANALYSIS

The SRBA is used to determine the content of optical solitons generated in our system for different initial conditions (Eq. 4, Eq. 5, and Eq. 6). It will provide us with information about solitons' structure and order[20,21].

We briefly explain the numerical technique of SRBA by taking the example when the propagation of the initial wave (Eq. 6) occurs for the input power of $P_0 = 3.3$ W. The total fibre length is set to $L = 20$ km. Generally, the resolution of the power spectrum plots within the SRBA strongly depends on the total fibre length chosen for simulation. Precisely, it goes with $1/L$ with $L$ being the total propagation length. Therefore, it is advisable to choose long fibre lengths to obtain sharp spectral lines.

To perform the SRBA, one first needs to calculate the optical field along the propagation distance, i.e. $A(z,t)$, then the optical power $P(z,t) = |A(z,t)|^2$. After that, the optical power at $t = 0$ is extracted, i.e. $P(z) = P(z,t=0)$. As presented in Fig. 3, the function $P(z)$ oscillates over the propagation distance (black curve)[39]. This oscillation contains information about the solitons involved and manifests itself in strong peaks in the power spectrum $\tilde{P}(Z)$ obtained via a Fourier transform of $P(z)$ (Fig. 4, black curve). To emphasise the spectral power peaks, we perform the Fourier transform on the apodised data that are depicted as the red curve in Fig. 3. The resulting spectral power is shown as the red curve in Fig. 4. The apodisation function is a Gaussian with the maximum positioned at the middle of the total propagation distance:

$$f(k) = \exp\left[-\left(\frac{k-K/2}{bK}\right)^2\right] \quad (8)$$

with $1/b$ being the apodisation strength and $k \in [1,\ldots,K]$, where K=20000 is the total number of distance sampling points. For further studies, $b$ is set to $b = 0.2$

## 4. RESULTS

Now, using the SRBA technique presented in Sec. 3, we analyse our system for input power values of $0.03\text{ W} \leq P_0 \leq 6.00\text{ W}$, the total length of fibre A is chosen to be $L = 20$ km. We plot the spectral power as a function of spatial frequency $Z$ and the input power $P_0$. In all graphs presented below, one will see a strong peak for any values of $P_0$ and $Z = 0\text{ km}^{-1}$. This peak arises during the Fourier transform from the optical into frequency domain and corresponds to the average value of the optical power $P(z)$. Since it does not contain any information about the soliton content, we will exclude it from consideration. Another feature of the SRBA is the appearance of the overtones of the oscillations. The overtones give us no further information about the soliton content. Therefore, they will be also excluded from consideration.

### 4.1 Single cosine-hump as initial condition

Fig. 5 shows the logarithmic power spectrum for different values of the input power obtained using a single cosine-hump as IC (Eq. 1 and Eq. 4). It is typical for single solitons to arise at a positive threshold value of the input power, $P_0 > 0$ W, and to evolve depending on $\sqrt{P_0}$[19,20]. In our case, such typical behaviour is presented by branch S1 that starts from $P_0 = 0.7$ W at $Z = 0$ km$^{-1}$. Thus, we can conclude that S1 constitutes a beating of a single soliton with the background. The scale order of this soliton is $N = 0.62$ according to Eq. 7. For the NLS, the threshold for creation of fundamental solitons is $N = 0.5$[19]. Thus, S1 can be identified as a fundamental soliton.

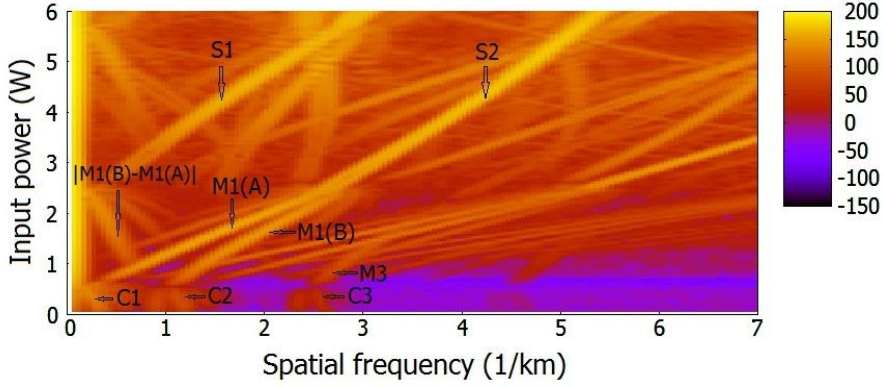

FIG. 6. Power spectrum for an Akhmediev breather as initial condition for different values of the input power

The next soliton branch S2 arises at $P_0 = 3.3$ W and $Z = 0$ km$^{-1}$ and has the scale soliton order $N = 1.35$. For the NLS, the threshold for a second-order soliton is at $N = 1.5$[19]. Since the scale order of the soliton involved into the branch S2 is below this threshold, S2 can be identified as another fundamental soliton. Generally, the spatial frequency scales with the energy of the solitons. The energy growth of S1 with increasing input power starts decreasing as soon as S2 appears. This has a change of the slope of S1 as a result meaning that the energy provided by the initial field is now split between both solitons.

The branches named as O1, O2, O3 are the overtones of S1.

### 4.2 Akhmediev breather as initial condition

Fig. 6 represents the power spectrum obtained by choosing a maximally compressed Akhmediev breather as IC (Eq. 1 and Eq. 5). Clearly, there are three input-power dependent regions with different soliton behaviour, i.e. $0.03$ W $\leq P_0 < 0.54$ W, $0.54$ W $< P_0 < 2.3$ W, and $P_0 > 2.3$ W.

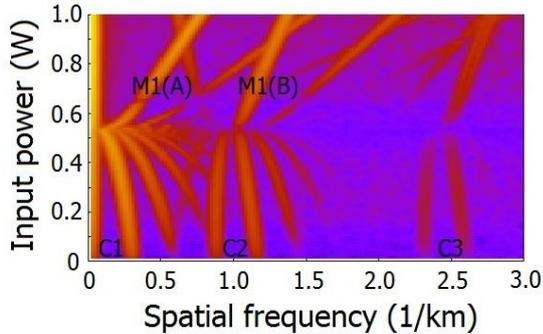

FIG. 7. Power spectrum for the Akhmediev initial condition at low input powers

For input powers $0.03$ W $\leq P_0 < 0.54$ W, we observe three branches C1, C2, C3 that are not well resolved. To obtain a more detailed presentation of C1, C2, and C3, we choose a fibre length of $L = 50$ km. The resulting power spectrum is depicted in Fig. 7. The three branches and their overtones are now presented in more detail. Obviously, there is no input-power threshold for the formation of the C-branches visible, i.e. C1, C2, and C3 arise for $P_0 = 0$ W at $Z = 0.3$ km$^{-1}$, $Z = 1.2$ km$^{-1}$, and $= 2.6$ km$^{-1}$, respectively. As discussed in Sec 4.1, a single cosine-hump as initial condition does not have enough energy to form a soliton at low input powers. However, as we have here a periodic initial condition, a collective soliton state can be formed even with less input power, creating the C-branches. The required energy is provided by the initial condition that incorporates $\cos-$functions that have infinite energy for $t \to \pm\infty$ (Eq. 5). By analogy to an electronic state in a crystal, the C-branches can be referred to as a collective soliton crystal state (cf. Ref. 40).

For input powers $P_0 > 0.54$ W, significant groups of branches arise out of C1, C2, C3. Branch M1(A) originates from C1, M1(B) from C2, and M3 from C3. Branches M1(A) and M1(B) merge for increasing value of $P_0$. To explain this behaviour, we start at the higher input-power edge of this region, i.e. at $P_0 = 2.3$ W. Beyond this region, i.e. $P_0 > 2.3$ W the soliton branches (S1 and S2) are temporally well separated since their duration is small compared to their temporal separation (cf. Ref. 41, Ref. 42). As the input power and, so, the solitons' energies decrease in the region $P_0 < 2.3$ W, the duration of solitons increases. Eventually, the solitons overlap in the temporal domain which makes solitons' energies split: M1(A) and M1(B) arise[19]. In analogy to the energy splitting in molecules, M1(A) and M1(B) can be regarded as a common soliton molecule state M1[43-48].

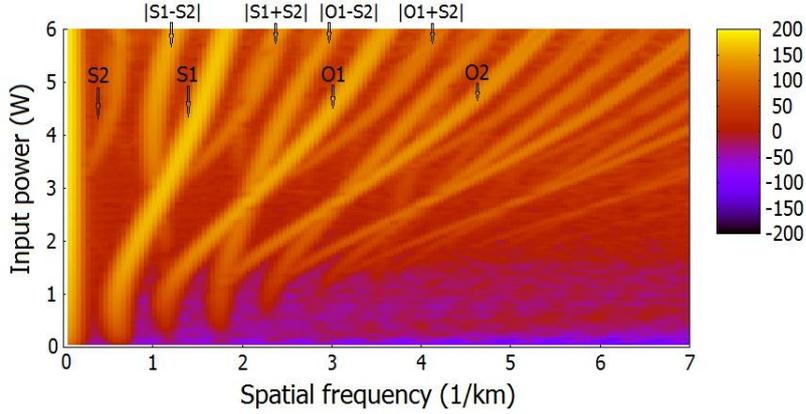

FIG. 8. Power spectrum for a deeply modulated cosine-wave as initial condition for different values of the input power

The branch $|M1(B) - M1(B)|$ constitutes the mixing frequency between M1(A) and M1(B).

For exactly $P_0 = 0.54$ W, we observe a perfect oscillation-free Akhmediev breather which manifests inself in a thin line that parts the collective soliton state from the state of soliton molecules in Fig. 7.

### 4.3 Modulated cosine-wave as initial condition

Fig. 8 presents the power spectrum obtained using a deeply modulated cosine-wave as IC (Eq. 6). The most prominent soliton branch S1 starts at $P_0 = 0$ W and $Z = 0.65$ km$^{-1}$. Similar to the case when Akhmediev breather was chosen as IC, the branch S1 has no input-power threshold constituting a collective soliton crystal state. The input power range where the collective state exists is $0 \text{ W} \leq P_0 < 1.3$ W. In case of an Akhmediev breather as IC, the region of the collective state was separated from the molecule state at $P_0 = 0.54$ W. Here, the transition from the soliton crystal to a state of well separated solitons occurs continuously which is indicated by a smooth evolution of the S1-branch for increasing input powers. A molecule state that manifests itself in the splitting of the lines was not observable.

At $P_0 = 3.3$ W and $Z = 0$ km$^{-1}$, a second soliton branch S2 arises. Obviously, it shows the behaviour of a single soliton beating with the background (cf. Sec. 4.1). According to Eq. 7, it has the soliton scale order of $N = 1.35$ meaning that S2 represents a fundamental soliton.

The overtones of S1 arise at $P_0 = 0$ W and $Z = 1.2$ km$^{-1}$ (O1) and $Z = 1.8$ km$^{-1}$ (O2). The branches $|S1 - S2|$, $|S1 + S2|$, $|O1 - S2|$, and $|O1 + S2|$ constitute the frequencies emerged from the beating of solitons S1 and S2 with each other or of solitons S and the overtones O.

### 5. CONCLUSION

The aim of this study was to understand the build-up of the temporal pulse shape in the first fibre stage of the proposed setup for generation of optical frequency combs for the calibration of astronomical spectrographs in the low- and medium resolution range. A bichromatic deeply modulated cosine-wave is chosen as optical input field within the framework of this setup. In particular, the temporal behaviour needs to be considered for two limiting cases: for high input powers and input powers that go to zero. The temporal build-up of a bichromatic cosine-wave in the first fibre stage is easily understood if compared with the cases when a single cosine-hump and a maximally compressed Akhmediev breather are chosen as initial condition.

It was expected that a single cosine-hump as initial condition will evolve into a soliton while propagating through the fibre[49]. This behaviour was confirmed in our studies: we observed the emergence of two fundamental solitons depending on the input power, the soliton S1 emerged for the input power of 0.7 W and the soliton S2 for 3.3 W. Comparing the relative soliton energy content at the limit of high input powers, we see that oscillation of the soliton S1 is faster than the soliton S2, namely, the spatial frequency of S1 is 1.8 km$^{-1}$ and the frequency of S2 is 0.9 km$^{-1}$ for the input power of 6.0 W.

The case when a maximally compressed Akhmediev breather is chosen as initial condition is more complicated than the previous one. Thus, we observe the build-up of an oscillation-free Akhmediev breather at the input power of 0.54 W. For lower input powers, we observe an oscillating collective soliton crystal state. For input powers $0.54 \text{ W} < P_0 < 2.3$ W, there is an intermediate soliton molecule state that is followed by the state of separated solitons as the input power increases. In the high input-power limit we observe two separated solitons, S1 and S2. S1 has less energy than S2 and oscillates slower. So, S1 has the spatial frequency of 3.0 km$^{-1}$ and S2 the frequency 5.3 km$^{-1}$ at the input power of 6.0 W. Besides these two solitons, there are many other oscillations that make the usage of an Akhmediev breather as initial condition in a real setup unsuitable, because these oscillations will affect the quality of optical frequency combs in terms of stability and noise evolution.

The case we are interested in to generate optical frequency combs for the purpose of calibration of astronomical low- and medium resolution spectrographs is given when a bichromatic deeply modulated cosine-wave is chosen as initial condition. It is quantitatively and qualitatively similar to the case when we have only a single cosine-hump as initial condition. That means there are two separated solitons S1 and S2 for the input power of 6.0 W, the faster soliton S1 has the spatial frequency 1.7 km$^{-1}$, the slower soliton S2 the frequency 0.6 km$^{-1}$. For the astronomical application, a multi-soliton state is not suitable because such states tend to soliton fission which might deteriorate the quality of optical frequency combs[18]. In the low input-power limit, the build-up of optical structures is similar to the case when an Akhmediev breather is chosen as initial condition. However, the input power is too low to give rise to a broadband optical frequency comb.

In the intermediate input-power region (2.0 W < $P_0$ < 3.0 W), there is state that suits well the requirements of the astronomical applications. Here, only a fundamental soliton S1 is generated. So, there is no danger of soliton fission. Further, the temporal and spatial periodicity coming from the soliton crystal state is still imprinted into the features of S1. The overall dynamics of this soliton are comparably simple and give rise to formation of a good-natured optical frequency comb. Thus, this input-power region should be chosen in a real experiment.